\documentclass[3p,times,twocolumn]{elsarticle}

\usepackage{ecrc}
\usepackage{widetext}

\volume{00}

\firstpage{1}

\journalname{Nuclear and Particle Physics Proceedings}

\runauth{}

\jid{nppp}

\jnltitlelogo{Nuclear and Particle Physics Proceedings}

\usepackage{amssymb}

\usepackage[figuresright]{rotating}

\begin{document}

\begin{frontmatter}



\dochead{}

\title{Subleading power corrections in radiative leptonic $B$ decay}


\author{Yu-Ming Wang}

\address{Fakult\"{a}t f\"{u}r Physik, Universit\"{a}t Wien, Boltzmanngasse 5, 1090 Vienna, Austria \\
School of Physics, Nankai University, 300071 Tianjin, China}

\begin{abstract}

I discuss the two-particle subleading power  corrections in  radiative leptonic $B \to \gamma \ell \nu$ decay
at next-to-leading order in $\alpha_s$ with the dispersion approach. Employing the method of regions,
factorization of the $B \to \gamma^{\ast}$ form factors is demonstrated explicitly, at one loop,
for a space-like hard-collinear photon. The two-particle soft (end-point) contribution is shown to be suppressed
by one power of $\Lambda/m_b$, in the heavy quark limit, compared with the leading power contribution computed from
QCD factorization. I further report the recent calculation on the three-particle subleading power contribution to the on-shell
$B \to \gamma$ form factors at tree level and demonstrate that the ``soft" and the ``hard" three-particle corrections
are of the  same power, in contrast to the two-particle counterparts, as already speculated from the rapidity divergence
in the corresponding factorization formulae. Phenomenological implications of the subleading power contributions to the
$B \to \gamma \ell \nu$ amplitude are also addressed in detail, focusing on the determination of the inverse moment
of the leading-twist $B$-meson distribution amplitude.

\end{abstract}


\end{frontmatter}


\section{Introduction}
\label{sec: introduction}

Understanding subleading power corrections in heavy quark decays is of interest to explore
the general properties of heavy-quark expansion and to perform a stringent test of the CKM mechanism
of the Standard Model. The radiative leptonic  $B \to \gamma \ell \nu$ decay process involving only a single hadron
is considered to be one of  benchmark channels to investigate the power suppressed contributions in exclusive
$B$-meson decays. At leading power in $\Lambda/m_b$, soft-collinear factorization properties of  $B \to \gamma \ell \nu$
have been explored in both QCD \cite{Korchemsky:1999qb,DescotesGenon:2002mw} and soft-collinear effective theory (SCET)
\cite{Lunghi:2002ju,Bosch:2003fc}. Subleading power contributions in the  $B \to \gamma \ell \nu$ amplitude
including both the local and non-local hadronic effects have been discussed in QCD factorization at tree level \cite{Beneke:2011nf},
where the non-local power correction from the hard-collinear quark propagator was found to
preserve the symmetry relations for the $B \to \gamma$ form factors due to the
helicity conservation in the heavy quark limit. Subsequently,  two-particle subleading power corrections to  the
$B \to \gamma$ form factors were computed from the dispersion approach at tree level \cite{Braun:2012kp},
following the technique developed in the context of $\gamma^{\ast} \gamma \to \pi$ form factor \cite{Khodjamirian:1997tk}.
I will discuss the soft two-particle contribution to $B \to \gamma \ell \nu$ at ${\cal O}(\alpha_s)$ and the
three-particle subleading power contribution at tree level, as computed in \cite{Wang:2016qii}, with the dispersion approach.

The presentation is organized as follows.  I will first outline the general strategy of applying the dispersion approach
in the radiative leptonic  $B \to \gamma \ell \nu$ decay and then demonstrate QCD factorization for the two-particle contribution
to the generalized $B \to \gamma^{\ast}$  form factors at one loop.
Afterwards the three-particle contribution to the $B \to \gamma \ell \nu$ amplitude will be  discussed at tree level.
Numerical impact of the newly computed power suppressed contributions on the  $B \to \gamma$ form factors
and on the extraction of the inverse moment $\lambda_B$ will be further presented
with two different models for the two-particle $B$-meson distribution amplitudes (DA).

\section{Dispersion relations for the radiative leptonic $B \to \gamma \ell \nu$ decay}
\label{sec: dispersion relations}

We will start with some general aspects of the $B \to \gamma \ell \nu$ decay amplitude following
the theory overview presented in \cite{Beneke:2011nf,Wang:2016qii}.
To the first order in the electromagnetic correction the transition amplitude
for the  $B \to \gamma \ell \nu$ decay can be expressed as
\begin{widetext}
\begin{eqnarray}
{\cal A}(B^{-} \to \gamma \, \ell \ \nu) =
 {G_F \, V_{ub} \over \sqrt{2}} \, \left ( i \, g_{em} \, \epsilon_{\nu}^{\ast}  \right )
\bigg \{ T^{\nu \mu}(p, q) \, \overline \ell \, \gamma_{\mu} \, (1-\gamma_5)  \nu
 + Q_{\ell} \,\, f_B \,\,\overline \ell \, \gamma^{\nu} \, (1-\gamma_5)  \nu  \bigg \} \,,
\label{original B to gamma l nu amplitude}
\end{eqnarray}
where the two terms in the bracket describe the photon radiation from the hadron constitutes
and the lepton, respectively, and the hadronic tensor $T^{\nu \mu}$ is defined as follows
\begin{eqnarray}
T_{\nu \mu}(p, q) &\equiv& \int d^4 x \, e^{i p \cdot x}  \,
\langle 0 | {\rm T} \{j_{\nu, \rm{em}}(x),
\left [\bar u \gamma_{\mu} (1-\gamma_5) b \right ] (0) \} |  B^{-}(p+q) \rangle \,.
\end{eqnarray}
\end{widetext}
Applying the electromagnetic Ward identity $p_{\nu} \, T^{\nu \mu}(p, q) = -(Q_b-Q_u) \, f_B \, p_B^{\mu}$
and redefining the axial-vector $B \to \gamma$ form factor to absorb the second term in the bracket of
(\ref{original B to gamma l nu amplitude}) lead to the replacement rule
\begin{widetext}
\begin{eqnarray}
T_{\nu \mu}(p, q) &\rightarrow& - i \, v \cdot p \, \epsilon_{\mu \nu \rho \sigma}
\, n^{\rho} \, v^{\sigma} \, F_V(n \cdot p)
+ \left [ g_{\mu \nu} \,v \cdot p - v_{\nu} \, p_{\mu} \right ] \,
\underbrace{\left [ \hat{F}_A(n \cdot p) +  \frac{Q_{\ell} \, f_B}{v \cdot p} \right ] } \nonumber \\
&&  -Q_{\ell} \, f_B \, g_{\mu \nu} \,, \hspace{6.0 cm}  \equiv F_A(n \cdot p)
\end{eqnarray}
from which the differential decay rate of $B \to \gamma \ell \nu$ in the rest frame of the $B$-meson
can be  computed as
\begin{eqnarray}
\frac{d \, \Gamma}{ d \, E_{\rm \gamma}} \left ( B \to \gamma \ell \nu \right )
=\frac{\alpha_{em}^2 \, G_F^2 \, |V_{ub}|^2}{6 \, \pi^2} \, m_B \, E_{\gamma}^3 \,
\left ( 1- \frac{2 \, E_{\gamma}}{m_B} \right ) \,
\left [ F_V^2(n \cdot p) + F_A^2(n \cdot p) \right ] \,.
\end{eqnarray}
\end{widetext}
To explain the essential technique of the dispersion approach for the evaluation of
the subleading power contributions, we start with the correlation function
describing the off-shell $B \to \gamma^{\ast}$  transition
with a space-like  hard-collinear (transverse polarized) photon
following the discussion in \cite{Wang:2016qii,Khodjamirian:2006st,DeFazio:2005dx,DeFazio:2007hw}
\begin{widetext}
\begin{eqnarray}
\tilde{T}_{\nu \mu}(p, q) &\equiv& \int d^4 x \, e^{i p \cdot x}  \,
\langle 0 | {\rm T} \{ j_{\nu, \rm{em}}^{\perp}(x),
\left [\bar u \gamma_{\mu \, \perp} (1-\gamma_5) b \right ] (0) \} |  B^{-}(p+q) \rangle \big|_{p^2<0} \,, \nonumber \\
&=&  v \cdot p \, \left [ - i \, \epsilon_{\mu \nu \rho \sigma}
\, n^{\rho} \, v^{\sigma} \, F_V^{B \to \gamma^{\ast}}(n \cdot p, \bar n \cdot p)
+  g_{\mu \nu}^{\perp}  \, \hat{F}_A^{B \to \gamma^{\ast}} (n \cdot p, \bar n \cdot p) \right ]  \,,
\label{def: correlation function}
\end{eqnarray}
with the power counting rule for the external momentum $n \cdot p  \sim {\cal O}(m_b) \,, |\bar n \cdot p| \sim {\cal O}(\Lambda)$.
Taking advantage of the analytical property of the generalized $B \to \gamma^{\ast}$ form factors
yields the hadronic dispersion relations \cite{Wang:2016qii}
\begin{eqnarray}
F_V^{B \to \gamma^{\ast}}(n \cdot p, \bar n \cdot p)
&=&{2 \over 3} \, \frac{f_{\rho} \, m_{\rho}}{m_{\rho}^2-p^2-i 0}
\, {2 \, m_B \over m_B +  m_{\rho}} \, V(q^2)  + {1 \over \pi} \, \int_{\omega_s}^{\infty} \, d \omega^{\prime} \,\,
\frac{{\rm Im}_{\omega^{\prime}} \, F_V^{B \to \gamma^{\ast}, \, {\rm had}}(n \cdot p, \omega^{\prime})}
{\omega^{\prime}- \bar n \cdot p - i 0}  \,, \,\,
\label{dispersion relation: FV}\\
\hat{F}_A^{B \to \gamma^{\ast}}(n \cdot p, \bar n \cdot p)
&=& {2 \over 3} \, \frac{f_{\rho} \, m_{\rho}}{m_{\rho}^2-p^2-i 0}
\, { 2 \left( m_B +  m_{\rho} \right ) \over n \cdot p } \, A_1(q^2)
 + {1 \over \pi} \, \int_{\omega_s}^{\infty} \, d \omega^{\prime} \,\,
\frac{{\rm Im}_{\omega^{\prime}} \, \hat{F}_A^{B \to \gamma^{\ast}, \, {\rm had}}(n \cdot p, \omega^{\prime})}
{\omega^{\prime}- \bar n \cdot p - i 0}  \, \,.
\label{dispersion relation: FAhat}
\end{eqnarray}
\end{widetext}
Applying the light-cone operator-product-expansion (OPE) technique and working out
the dispersion representations for the  resulting factorization formulae of
the $B \to \gamma^{\ast}$ form factors lead to the light-cone sum rules for the
form factors $V(q^2)$ and $A_1(q^2)$
\begin{widetext}
\begin{eqnarray}
{2 \over 3} \, \frac{f_{\rho} \, m_{\rho}}{n \cdot p} \,
{\rm Exp} \left [-{m_{\rho}^2 \over n \cdot p \, \omega_M} \right ] \,
{2 \, m_B \over m_B +  m_{\rho}} \, V(q^2)
&=&  {1 \over \pi} \, \int_0^{\omega_s} \,\, d \omega^{\prime} \,\,  e^{-\omega^{\prime}/\omega_M} \,
\, {\rm Im}_{\omega^{\prime}} \, F_V^{B \to \gamma^{\ast}}(n \cdot p, \omega^{\prime})  \,, \hspace{0.8 cm}
\label{sum rules of the form factor V} \\
{2 \over 3} \, \frac{f_{\rho} \, m_{\rho}}{n \cdot p} \,
{\rm Exp} \left [-{m_{\rho}^2 \over n \cdot p \, \omega_M} \right ] \,
 { 2 \left( m_B +  m_{\rho} \right ) \over n \cdot p } \, A_1(q^2)
&=& {1 \over \pi} \, \int_0^{\omega_s} \,\, d \omega^{\prime} \,\,  e^{-\omega^{\prime}/\omega_M} \,
\,{\rm Im}_{\omega^{\prime}} \, \hat{F}_A^{B \to \gamma^{\ast}}(n \cdot p, \omega^{\prime})  \,.
\label{sum rules of the form factor A1}
\end{eqnarray}
\end{widetext}
Substituting the above sum rules into (\ref{dispersion relation: FV}) and  (\ref{dispersion relation: FAhat})
and setting $\bar n \cdot p \to 0$ give rise to improved dispersion relations for the on-shell $B \to \gamma$
form factors
\begin{widetext}
\begin{eqnarray}
F_V(n \cdot p) &=& {1 \over \pi} \, \int_0^{\omega_s} \,\, d \omega^{\prime} \,\,  \frac{n \cdot p}{m_{\rho}^2} \,
{\rm Exp} \left [{m_{\rho}^2 - \omega^{\prime} \, n \cdot p \over n \cdot p \, \omega_M} \right ]
\, \left [{\rm Im}_{\omega^{\prime}} \, F_V^{B \to \gamma^{\ast}}(n \cdot p, \omega^{\prime}) \right ]  \,  \nonumber \\
&& +  {1 \over \pi} \, \int_{\omega_s}^{\infty} \,\, d \omega^{\prime} \,\,  \frac{1}{\omega^{\prime}} \,
\, \left [{\rm Im}_{\omega^{\prime}} \, F_V^{B \to \gamma^{\ast}}(n \cdot p, \omega^{\prime}) \right ] \,,
\label{master formula of FV} \\
\hat{F}_A(n \cdot p) &=& {1 \over \pi} \, \int_0^{\omega_s} \,\, d \omega^{\prime} \,\,  \frac{n \cdot p}{m_{\rho}^2} \,
{\rm Exp} \left [{m_{\rho}^2 - \omega^{\prime} \, n \cdot p \over n \cdot p \, \omega_M} \right ]
\, \left [{\rm Im}_{\omega^{\prime}} \, \hat{F}_A^{B \to \gamma^{\ast}}(n \cdot p, \omega^{\prime}) \right ]  \,  \nonumber \\
&& +  {1 \over \pi} \, \int_{\omega_s}^{\infty} \,\, d \omega^{\prime} \,\,  \frac{1}{\omega^{\prime}} \,
\, \left [{\rm Im}_{\omega^{\prime}} \, \hat{F}_A^{B \to \gamma^{\ast}}(n \cdot p, \omega^{\prime}) \right ] \,,
\label{master formula of FAhat}
\end{eqnarray}
where the second term on the right-hand side of (\ref{master formula of FV}) and (\ref{master formula of FAhat})
corresponds to the soft (end-point) contribution due to the nonpertubative modification of the  spectral density.
\end{widetext}

\begin{widetext}
At tree level, the generalized $B \to \gamma^{\ast}$ form factors can be readily computed as
\begin{eqnarray}
F_{V, \,2P}^{B \to \gamma^{\ast}}(n \cdot p, \bar n \cdot p)
=\hat{F}_{A, \,2P}^{B \to \gamma^{\ast}}(n \cdot p, \bar n \cdot p)
= \frac{Q_u \, \tilde{f}_B(\mu) \, m_B}{n \cdot p} \,
\, \int_0^{\infty} \, d \omega \, \frac{\phi_B^{+}(\omega, \mu)}{\omega - \bar n \cdot p - i 0} \,
+ \,\, {\cal O}(\alpha_s, \Lambda/m_b) \,.
\label{correlator: QCD at tree level}
\end{eqnarray}
Now we turn to discuss the computation of the one-loop hard and jet functions in the factorization formulae
\begin{eqnarray}
&& F_{V}^{B \to \gamma^{\ast}}(n \cdot p, \bar n \cdot p)
= \hat{F}_{A}^{B \to \gamma^{\ast}}(n \cdot p, \bar n \cdot p)  \nonumber  \\
&& = \frac{Q_u \, \tilde{f}_B(\mu) \, m_B}{n \cdot p} \, C_{\perp}(n \cdot p, \mu)  \,
\, \int_0^{\infty} \, d \omega \, \frac{\phi_B^{+}(\omega, \mu)}{\omega - \bar n \cdot p - i 0} \,
J_{\perp}(n \cdot p, \bar n \cdot p , \omega, \mu) + ... \,,
\label{master formulae for one-loop two particle factorization formula}
\end{eqnarray}
at leading power in $\Lambda/m_b$, employing the method of regions \cite{Beneke:1997zp}
which has been extensively used for evaluating the multi-scale amplitudes
(see, for instance \cite{Beneke:2004rc,Beneke:2005gs,Wang:2015vgv,Wang:2015ndk,Wang:2016slj}).
\end{widetext}

\begin{figure}[t]
\begin{center}
\includegraphics[width=8 cm]{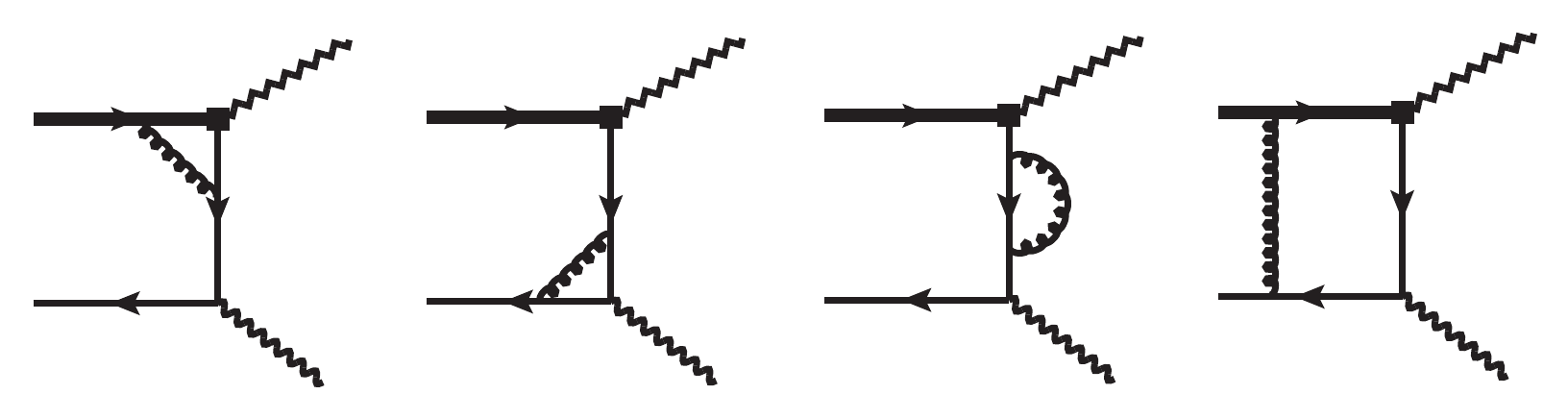} \\
(a) \hspace{1.5 cm} (b) \hspace{1.5 cm}  (c) \hspace{1.5 cm}   (d)
\end{center}
\vspace*{-0.3cm}
\caption{Two-particle contribution to the correlation function (\ref{def: correlation function})
at one loop. }
\label{loop diagrams of the correlator}
\end{figure}

Taking the weak vertex diagram displayed in figure 1(a) as an example,
the corresponding QCD amplitude without the ultraviolet and infrared subtraction
can be readily computed as
\begin{widetext}
\begin{eqnarray}
\tilde{T}^{(1)}_{\nu \mu \,, weak}(p,q)
&=& \frac{Q_u \, g_s^2 \, C_F}{\bar n \cdot p -\omega} \, \int \frac{d^D l}{(2 \pi)^D} \,
\frac{1}{\left [(p-k+l)^2+ i 0 \right ] \, \left [(m_b \, v + l)^2 - m_b^2 + i 0 \right ]
\, \left [l^2 + i 0 \right ] } \nonumber \\
&& \left \{ n \cdot l  \left [ (D-2) \, \bar n \cdot l + 2\, m_b  \right ]   +  (D-4) \, l_{\perp}^2
+ 2\, n \cdot p \, (\bar n \cdot l + m_b)  \right \}
\bar u(k) \, \gamma_{\nu \perp} \, \frac{\not \! \bar n}{2} \, \gamma_{\mu \perp} \, \,(1 - \gamma_5) \, b(v)\,,
\label{full result of the weak vertex diagram}
\end{eqnarray}
\end{widetext}
Employing the power counting scheme for the external momenta $p$ and $k$, one can conclude that the leading
power contributions to the weak vertex diagram come from the hard, hard-collinear and soft regions.
It is evident that the soft contribution will be cancelled exactly by the corresponding
infrared subtraction term which is defined by the convolution integral of the
partonic distribution amplitude at next-to-leading-order (NLO) in $\alpha_s$ and the leading order (LO) hard kernel.
The resulting contribution to the hard coefficient function
$C_{\perp}$ can be extracted from (\ref{full result of the weak vertex diagram}) with the integrand
of the loop-momentum integral expanded in the hard region  \cite{Wang:2016qii}
\begin{widetext}
\begin{eqnarray}
\tilde{T}^{(1),\,  h}_{\nu \mu \,, weak}(p,q)
&=& - i\, g_s^2 \, C_F \, \int \frac{d^D l}{(2 \pi)^D} \,
\frac{\tilde{T}^{(0)}_{\nu \mu}(p,q)}{\left [l^2 + n \cdot p \, \bar n \cdot l + i 0 \right ] \,
\left [l^2 + 2 \, m_b \, v \cdot l + i 0 \right ]
\, \left [l^2 + i 0 \right ] }   \nonumber \\
&& \times \left \{ n \cdot l  \left [ (D-2) \, \bar n \cdot l + 2\, m_b  \right ]   +  (D-4) \, l_{\perp}^2
+ 2\, n \cdot p \, (\bar n \cdot l + m_b)  \right \}  \,  \, \nonumber  \\
&\equiv& C_{\perp, weak}(n \cdot p) \,\,\,  \tilde{T}^{(0)}_{\nu \mu}(p,q) \,,
\end{eqnarray}
where $\tilde{T}^{(0)}_{\nu \mu}$ is the tree-level contribution to the correlation function
(\ref{def: correlation function}) and $C_{\perp, weak}$ can be found in Eq. (3.7) of \cite{Wang:2016qii}.

\end{widetext}

Similarly,  the leading-power hard-collinear contribution to the weak vertex diagram can be obtained
by expanding (\ref{full result of the weak vertex diagram}) in the hard-collinear region
\begin{eqnarray}
&& \hspace{-1 cm} \tilde{T}^{(1),\,  hc}_{\nu \mu \,, weak}(p,q)  = - i\, g_s^2 \, C_F \, \int \frac{d^D l}{(2 \pi)^D} \, \nonumber \\
&& \hspace{-1 cm} \frac{2 \, m_b \, n \cdot (p+l) \,\, \tilde{T}^{(0)}_{\nu \mu}(p,q) }
{[ n \cdot (p+l) \,\bar n \cdot (p-k+l) + l_{\perp}^2  + i 0][ m_b \, n \cdot l+ i 0] [l^2+i0] }  \,\, \nonumber \\
&& \hspace{-1 cm} \equiv  J_{\perp, weak}(n \cdot p, \bar n \cdot p, \omega) \,\,\,  \tilde{T}^{(0)}_{\nu \mu}(p,q) \,,
\end{eqnarray}
where the explicit expression of $J_{\perp, weak}$ can be found in Eq. (3.9) of \cite{Wang:2016qii}.

Evaluating the leading power contributions to remaining diagrams with the same technique leads to
\begin{widetext}
\begin{eqnarray}
\hspace{-0.5 cm} C_{\perp}
&=& 1- \frac{\alpha_s \, C_F}{4 \, \pi}
\bigg [ 2 \, \ln^2 {\mu \over n \cdot p} + 5 \, \ln {\mu \over m_b}
-2 \, {\rm Li}_2 \left ( 1-{1 \over r} \right )  - \ln^2  r
+ \,  {3 r -2 \over 1 -r}  \, \ln r + {\pi^2 \over 12} + 6  \bigg ] \,,
\label{one loop hard function}  \\
\hspace{-0.5 cm} J_{\perp} &=& 1 + {\alpha_s \, C_F \over 4 \, \pi} \,
\bigg \{ \ln^2 { \mu^2 \over n \cdot p \,  (\omega - \bar n \cdot p)}  - {\pi^2 \over 6} - 1
- {\bar n \cdot p \over \omega} \, \ln {\bar n \cdot p - \omega \over \bar n \cdot p } \,
\left [ \ln { \mu^2 \over -p^2} + \ln { \mu^2 \over n \cdot p \,  (\omega - \bar n \cdot p)} + 3 \right ]  \bigg \}  \,.
\label{one loop jet function}
\end{eqnarray}
\end{widetext}
Resummation of the parametrically large logarithms in the hard function $C_{\perp}$ and
in the $B$-meson decay constant $\tilde{f}_B$ can be achieved by solving the following evolution equations
\begin{widetext}
\begin{eqnarray}
\frac{d C_{\perp}(n \cdot p, \mu)}{d \ln \mu} \,
= \left [ - \Gamma_{\rm {cusp}}(\alpha_s) \, \ln{\mu \over n \cdot p}
+ \gamma(\alpha_s)  \right ] \, C_{\perp}(n \cdot p, \mu) \,,  \qquad
\frac{d \tilde{f}_B(\mu)}{d \ln \mu} \,
= \tilde{\gamma}(\alpha_s) \, \tilde{f}_B(\mu) \,,
\end{eqnarray}
\end{widetext}
It is then straightforward to write down the resummation improved factorization formulae for
the $B \to \gamma^{\ast}$ form factors, from which one can derive the following dispersion relations for the
two-particle contributions with  the aid of (\ref{master formula of FV}) and (\ref{master formula of FAhat})
\begin{widetext}
\begin{eqnarray}
&& \hspace {-0.2 cm} F_{V, 2P}(n \cdot p)
= \hat{F}_{A,2P}(n \cdot p)  = \frac{Q_u \, m_B}{n \cdot p} \, \left [  U_2(n \cdot p, \mu_{h2}, \mu) \,\tilde{f}_B(\mu_{h2}) \right ] \,
\left [ U_1(n \cdot p, \mu_{h1}, \mu) \, C_{\perp}(n \cdot p, \mu_{h1})   \right ] \nonumber \\
&& \hspace{0.3 cm} \times \, \bigg \{   \,\int_0^{\infty} \, d \omega \,
 \frac{\phi_B^{+}(\omega, \mu)}{\omega} \,
J_{\perp}(n \cdot p, 0, \omega, \mu)  + \int_0^{\omega_s} \,\,d \omega^{\prime} \,\,  \left [ \frac{n \cdot p}{m_{\rho}^2} \,
{\rm Exp} \left [{m_{\rho}^2 - \omega^{\prime} \, n \cdot p \over n \cdot p \, \omega_M} \right ]
- {1 \over \omega^{\prime}} \right ] \, \phi_{B, {\rm eff}}^{+}(\omega^{\prime},\mu) \, \bigg \}  \,,
\label{NLL 2-particle contribution to form factors}
\end{eqnarray}
where the explicit expression of $\phi_{B, {\rm eff}}^{+}(\omega^{\prime},\mu) $ is displayed in Eq. (3.31) of \cite{Wang:2016qii}.
\end{widetext}


Now we turn to discuss the three-particle contribution to the generalized $B \to \gamma^{\ast}$
form factors at tree level, which can be obtained by evaluating the partonic diagram presented
in figure 3 of \cite{Wang:2016qii}. 
Applying the background field approach for the light-quark propagator we obtain

\begin{eqnarray}
&& \hspace{-0.2 cm} F_{V, \, 3P}^{B \to \gamma^{\ast}}(n \cdot p, \bar n \cdot p)
= \hat{F}_{A, \, 3P}^{B \to \gamma^{\ast}}(n \cdot p, \bar n \cdot p)  \nonumber  \\
&& \hspace{-0.2 cm} = - \frac{Q_u \, \tilde{f}_B(\mu) \, m_B}{(n \cdot p)^2} \,
\int_0^{\infty} d \omega \, \int_0^{\infty} d \xi \, \int_0^1 d u \, \nonumber  \\
&& \hspace{-0.2 cm} \bigg \{\frac{ \rho_{3P}^{(2)}(u,\omega,\xi)}{[\bar n \cdot p - \omega - u \, \xi]^2}
+ \frac{\rho_{3P}^{(3)}(u,\omega,\xi)}{[\bar n \cdot p - \omega - u \, \xi]^3} \bigg \} \,,
\label{3-particle contribution to B to gamma-star FFs}
\end{eqnarray}
where the manifest expressions of the spectral functions are shown in Eq. (4.4) of \cite{Wang:2016qii}.
Substituting (\ref{3-particle contribution to B to gamma-star FFs}) into the master formulae
(\ref{master formula of FV}) and (\ref{master formula of FAhat}) yields \cite{Wang:2016qii}
\begin{widetext}
\begin{eqnarray}
 F_{V, \, 3P}(n \cdot p)
= \hat{F}_{A, \, 3P}(n \cdot p)  = - \frac{Q_u \, \tilde{f}_B(\mu) \, m_B}{(n \cdot p)^2} \,
\left \{  \frac{n \cdot p}{m_{\rho}^2} \,
{\rm Exp} \left [{m_{\rho}^2 \, \over n \cdot p \, \omega_M} \right ] \,
I_{3P}^{\rm I}(\omega_s, \omega_M) +  I_{3P}^{\rm II}(\omega_s, \omega_M) \right \} \,,
\label{3-particle contribution to B to gamma FFs: dispersion approach}
\end{eqnarray}
where the first and second terms in the bracket correspond to the ``soft" and ``hard"
three-particle contributions. One can conclude from the power counting rules for the
external momenta and the sum-rule parameters that both the ``soft" and ``hard"
three-particle contributions scale as $(\Lambda/m_b)^{3/2}$ in the heavy quark limit.
\end{widetext}

To evaluate the numerical impact of the subleading power two-particle correction at ${\cal O}(\alpha_s)$
and the three-particle contribution at tree level, we adopt the two different models for the two-particle $B$-meson
DA inspired from the QCD sum rule analysis at LO and at NLO (see \cite{Feldmann:2014ika} for an improvement
including perturbative constraints) and employ the exponential model for the three-particle
$B$-meson DA. The key  quantity entering the parametrization of the above-mentioned nonperturbative functions
is the inverse moment $\lambda_B$ of the leading-twist $B$-meson DA, which also serves as a fundamental hadronic input
for the theoretical description of many other exclusive processes
\cite{Beneke:2000wa,Beneke:2001at,Khodjamirian:2010vf,Li:2012nk,Li:2012md}.
We will take the interval $\lambda_B(\mu_0)=354^{+38}_{-30} \, {\rm MeV}$  determined from the matching of
the two different types of  light-cone sum rules for the $B \to \pi$ form factors with the pion DA \cite{Khodjamirian:2011ub}
and with the $B$-meson DA \cite{Wang:2015vgv}, respectively.
With the default theory inputs, perturbative QCD corrections to the two-particle soft contribution is found to shift the
tree-level prediction by an amount of $(10 \sim 20) \%$, and the LO three-particle correction to the $B \to \gamma$
form factors turns out to be negligible numerically \cite{Wang:2016qii}.
However, the subleading power two-particle contribution can be enhanced significantly with the decrease of $\lambda_B$
and it is even comparable to the leading power contribution computed in QCD factorization at $\lambda_B \leq 100 \, {\rm MeV}$
as implied by the power counting analysis  \cite{Wang:2016qii}. Moreover,  the model dependence of the two-particle $B$-meson
DA on the theoretical predictions of $F_{V}$ and $F_{A}$ also becomes more important at small $\lambda_B$ and at small $E_{\gamma}$.
Finally, we are in a position to discuss the determination of the inverse moment $\lambda_B$ from the Belle
measurement of the integrated  branching ratio of $B \to \gamma \ell \nu$ \cite{Heller:2015vvm}.
Taking into account the newly computed subleading power corrections, no interesting constraint on $\lambda_B$
can be obtained for the Grozin-Neubert model \cite{Grozin:1996pq} due to the rather weak experiment limit,
while a meaningful bound $\lambda_B > 214 \, {\rm MeV}$ can be deduced for the Braun-Ivanov-Korchemsky model
\cite{Braun:2003wx} of the leading-twist $B$-meson DA.
This fact can be easily understood from the strong sensitivity of the $B \to \gamma$ form factors on the precise
shape of the $B$-meson DA $\phi_B^{+}(\omega)$ at small light-quark momentum $\omega$.


\section{Conclusions}
\label{sec: conclusions}

Applying the dispersion approach, the subleading power two-particle soft correction
to the $B \to \gamma \ell \nu$ transition amplitude was shown to be sizeable in particular at small $\lambda_B$
and the inverse moment $\lambda_B$ is not sufficient to describe the strong interaction dynamics of the
$B \to \gamma$ form factors in general. In contrast, the tree-level three-particle contribution can only lead to
the negligible impact on the $B \to \gamma$ form factors. Further improvement including  perturbative
QCD corrections to  the three-particle DA and the yet higher-twist corrections will be crucial to deepen
our understanding of the factorizaton properties in the heavy quark system and to achieve  precision determinations
of the CKM matrix elements.




\nocite{*}
\bibliographystyle{elsarticle-num}



\end{document}